\documentclass[useAMS,usenatbib]{mn2e}
\pdfoutput=1
\usepackage{graphicx}
\usepackage{epstopdf}
\usepackage{comment}
\usepackage{subfigure}
\title[A panoramic view of Fornax]{The shell game: a panoramic view of Fornax}
\author[N. F. Bate et al.]{N. F. Bate$^{1,2}$\thanks{E-mail:
nbate@ast.cam.ac.uk (NFB)}, B. McMonigal$^{1}$, G. F. Lewis$^{1}$, M. J. Irwin$^{3}$, E. Gonzalez-Solares$^{3}$, \newauthor T. Shanks$^{4}$ and N. Metcalfe$^{4}$ \\
$^{1}$Sydney Institute for Astronomy, School of Physics, A28, The University of Sydney, Sydney, NSW 2006, Australia\\
$^{2}$School of Physics, The University of Melbourne, Parkville, VIC 3010, Australia\\
$^{3}$Institute of Astronomy, University of Cambridge, Madingley Road, Cambridge CB3 0HA, UK\\
$^{4}$Department of Physics, Durham University, South Road, Durham, DH1 3LE, UK\\}
\begin{document}

\date{Accepted 2015 July 22. Received 2015 July 21; in original form 2015 May 29}

\pagerange{\pageref{firstpage}--\pageref{lastpage}} \pubyear{2015}

\maketitle

\label{firstpage}

\begin{abstract}
We present a panoramic study of the Fornax dwarf spheroidal galaxy, using data obtained as part of the VLT Survey Telescope (VST) ATLAS Survey. The data presented here -- a subset of the full survey -- uniformly cover a region of 25 square degrees centred on the galaxy, in $g$, $r$ and $i$-bands. This large area coverage reveals two key differences to previous studies of Fornax. First, data extending beyond the nominal tidal radius of the dwarf highlight the presence of a second distinct red giant branch population. This bluer red giant branch appears to be coeval with the horizontal branch population. Second, a shell structure located approximately 1.4 degrees from the centre of Fornax is shown to be a mis-identified background overdensity of galaxies. This last result casts further doubt on the hypothesis that Fornax underwent a gas-rich merger in its relatively recent past.
\end{abstract}


\section{Introduction}

\begin{figure*}	
  \includegraphics[width=140mm]{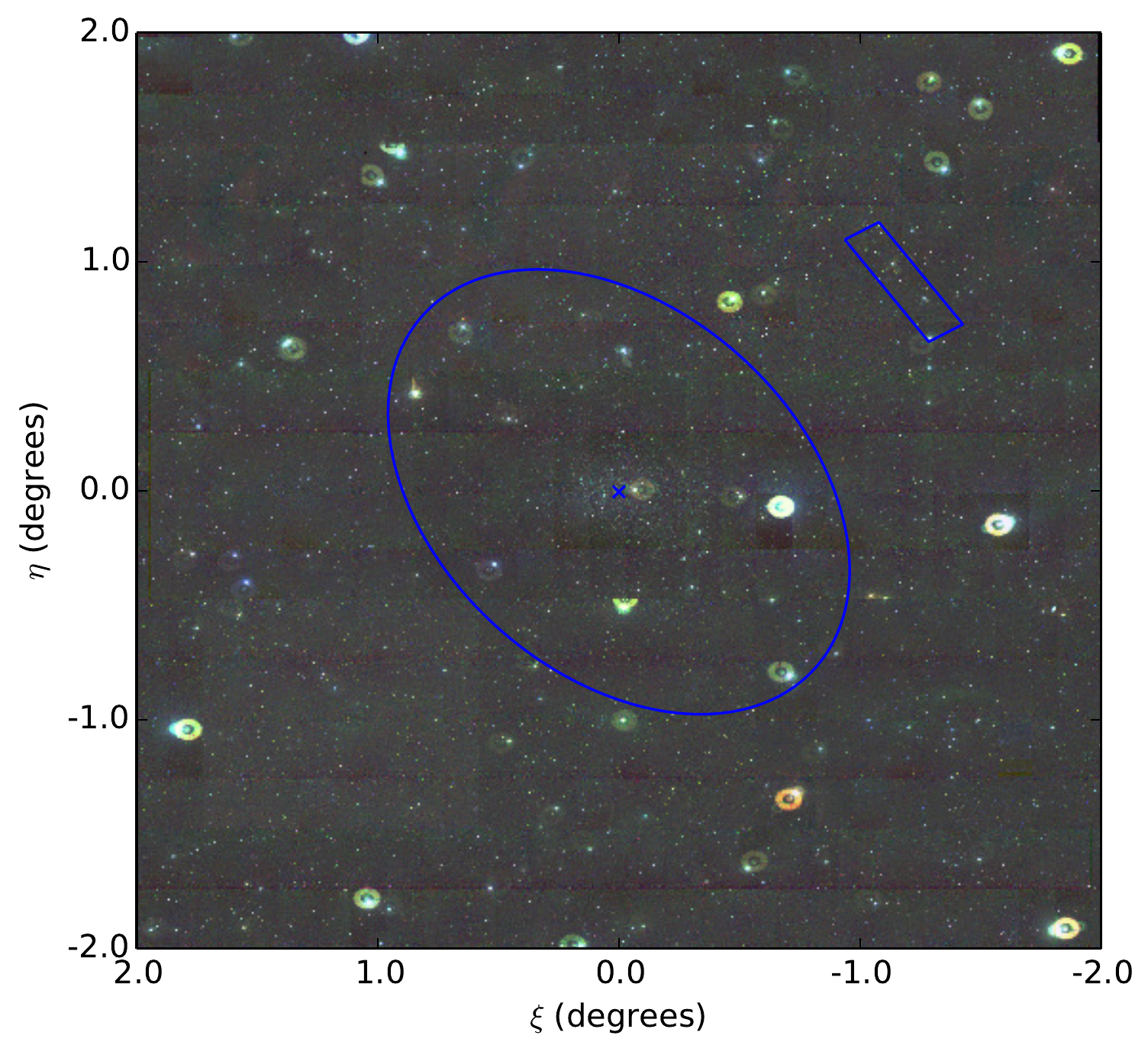}
  \caption{Three-colour image of 16 square degrees centred on the Fornax dwarf spheroidal, built from VST ATLAS $g$, $r$ and $i$-band imaging. The blue cross marks the centre of Fornax, and the blue ellipse marks the tidal radius \citep{Battaglia2006}. The blue box to the North-West marks the location of the outer shell identified in \citet{Coleman2005}. This image was generated using 23 separate pointings from the VST ATLAS Survey.}
  \label{fig:VST_image}
\end{figure*}

According to the standard $\Lambda$CDM cosmology, the Universe as we see it today is built via a scale-free process in which matter collapses under gravity and iteratively forms larger and larger structures. Dark matter dominates this hierarchical structure formation process. Dwarf galaxies, the most dark matter dominated objects known, are therefore crucial laboratories for studying galaxy formation and evolution processes. Considerable work has been invested in the study of dwarf galaxies in the Local Group, summarised recently in \citet{Tolstoy2009} and \citet{McConnachie2012}.

Local Group dwarf galaxies offer us the opportunity to examine resolved stellar populations in considerable detail. Such analyses have revealed richly complex star formation histories (e.g. \citealt{Bono2010}; \citealt{Battaglia2011}; \citealt{DeBoer2012a}, to name just three recent examples), and allow us to search for evidence for tidal interactions and mergers. This last point is particularly crucial, since tidal disturbance on shallow potentials can directly impact on the interpretation of dwarf galaxy properties. Wide-field imaging from instruments such as the Dark Energy Camera (DECam, e.g. \citealt{McMonigal2014}; \citealt{Roderick2015}) and survey telescopes such as the VLT Survey Telescope (VST) are making it possible to search for low surface brightness features in the outskirts of nearby dwarf galaxies.

The Fornax dwarf spheroidal galaxy is one of the more intensively studied satellites of the Milky Way. First discovered by \citet{Shapley38}, it is relatively distant at $147\pm12$kpc (all parameters in this paragraph from the review of \citealt{McConnachie2012}). Nevertheless, it is second only to the Sagittarius dwarf spheroidal in terms of luminosity ($M_V = -13.4\pm0.3$) and dynamical mass ($M_\rmn{dyn} = 5.6\times10^7M_\odot$ within the half light radius) amongst the MW dwarf population. For detailed information on the study of stellar populations in Fornax, we refer the reader to \citet{Battaglia2006}, \citet{DeBoer2012}, and \citet{DelPino2013}.

Fornax is of particular interest here because of a detection of a possible large outer shell, located approximately $1.4\degr$ from the centre of the dwarf \citep{Coleman2005}. The association of this shell with Fornax was circumstantially supported by an earlier detection of a smaller, inner feature with a shell-like morphology (hereafter referred to as clump 1, or C1; \citealt{Coleman2004}). Shells are found in large galaxies due to events such as mergers, and since hierarchical structure formation is scale free, we should expect to see them in dwarf galaxies too. So far, Fornax is the only dwarf galaxy reported to display unambiguous evidence of a shell, although kinematic measurements of the Andromeda II dwarf galaxy \citep{Amorisco2014} suggest that it too may have undergone a relatively major merger.

The initial interpretation of these two shells in Fornax was that the galaxy had undergone a gas-rich merger with another dwarf approximately 2 Gyr ago (\citealt{Coleman2004}; \citealt{Coleman2005}). However, this hypothesis has proven difficult to reconcile with the dwarf galaxy's star formation history, which shows a clear peak at intermediate ages; either 4 Gyr ago \citep{DeBoer2012}, or 8 Gyr ago \citep{DelPino2013}. 

Furthermore, \citet{Olszewski2006} demonstrated that the metallicity of the stars in C1 closely matched that of nearby field stars in Fornax. This result was confirmed by \citet{DeBoer2013}, who also detected a second overdensity in blue plume stars outside the centre of Fornax. This feature is composed of very young stars ($\sim100$ Myr old) with high metallicity, so is unlikely to have been formed from accreted pristine gas. 

A plausible explanation for these features, put forward by \citet{Olszewski2006}, \citet{Coleman2008} and \citet{Salvadori2008}, is that these clumps were formed from enriched gas blown out of Fornax and subsequently re-accreted. Such gas would fall back into the dwarf over a timescale of $\sim250$ Myr, sinking deeper as it is re-accreted. This is consistent with the radial stellar population age gradient observed by \citet{Battaglia2006}, suggesting the gradual removal of gas from Fornax, however the timescales are difficult to reconcile with its complex star formation history. Disrupted clusters can also produce shells or clumps \citep{Penarrubia2009}, however none of the globular clusters in Fornax show any signs of disruption (\citealt{Mackey2003}; \citealt{Rodgers1994}).

Tidal interactions with the Milky Way are not expected to have played a significant role in the formation of any of these stellar substructures. After Sagittarius, Fornax is the Milky Way's most massive dwarf spheroidal satellite. It is currently thought to be at or near perigalacticon, a large distance ($\sim150$ kpc, \citealt{McConnachie2012}) from the centre of the Galaxy, on a low-eccentricity polar orbit (e.g. \citealt{Dinescu2004}; \citealt{Piatek2007}). There is, however, some conflict in proper motion measurements. \citet{Dinescu2004} results indicate that Fornax may have had an encounter with the Magellanic Stream $\sim190$ Myr ago, which could explain very recent bursts of star formation, and is supported by small scale structure found along the proposed orbit. However, these results are not consistent with other proper motion measurements (e.g. \citealt{Piatek2002}).

\begin{figure}
\begin{center}
  \subfigure[Three colour VST image of the outer shell region]{
    \label{fig:zoom_a}	
    \includegraphics[width=80mm]{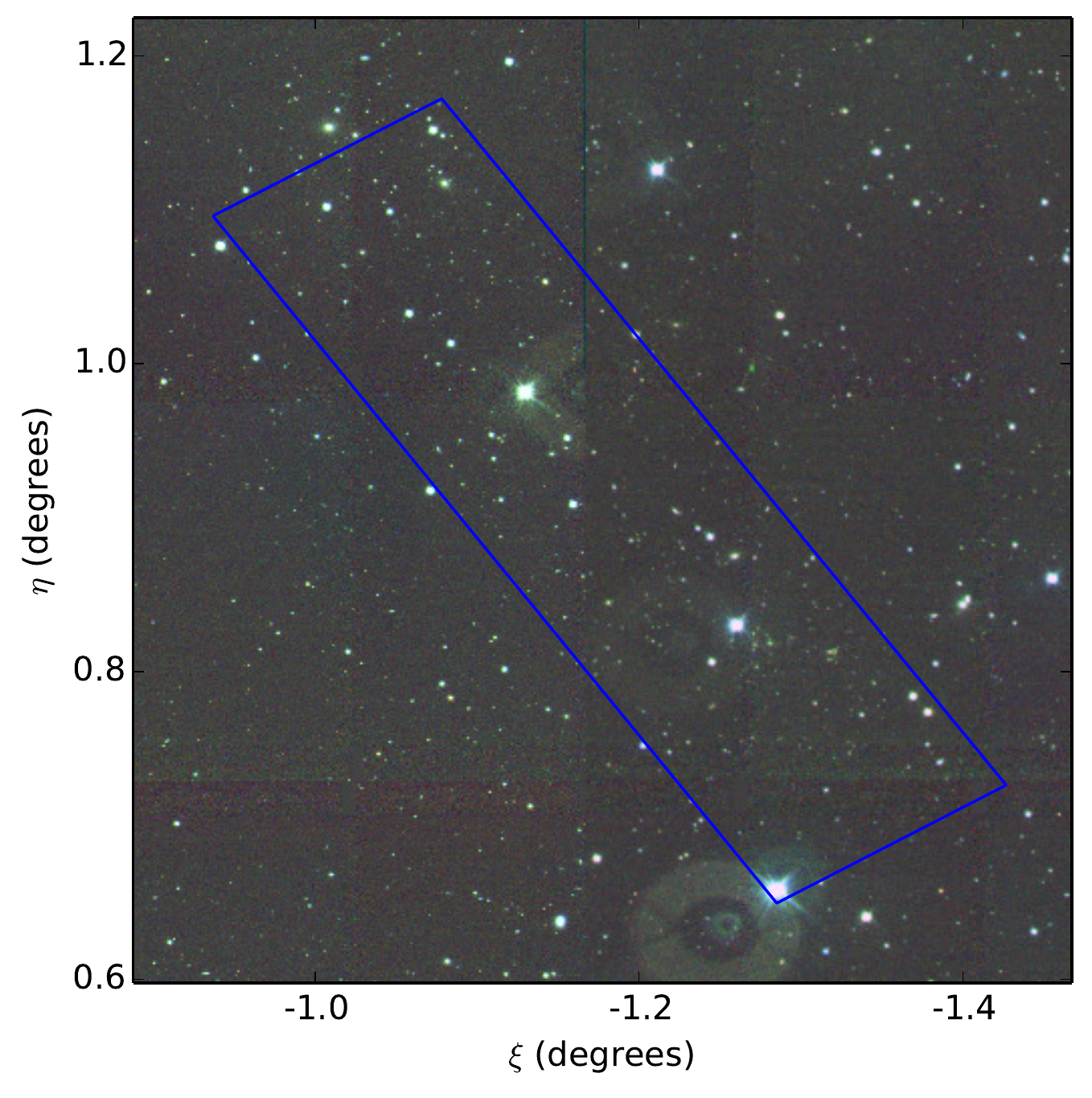}
  }\\
  \subfigure[Shell field]{
    \label{fig:zoom_b}
    \includegraphics[width=80mm]{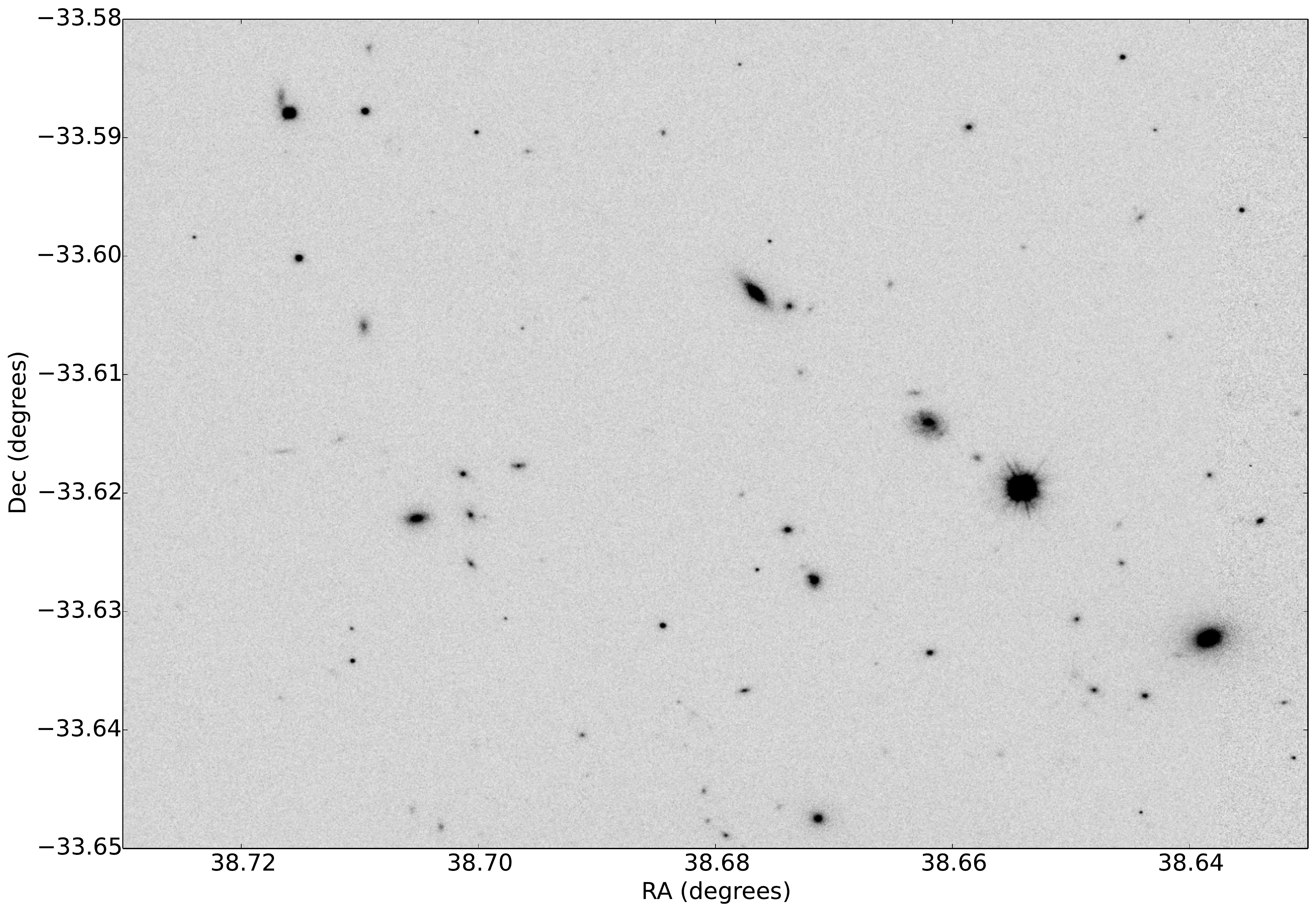}
  }\\
  \subfigure[Central Fornax field]{
    \label{fig:zoom_c}
    \includegraphics[width=80mm]{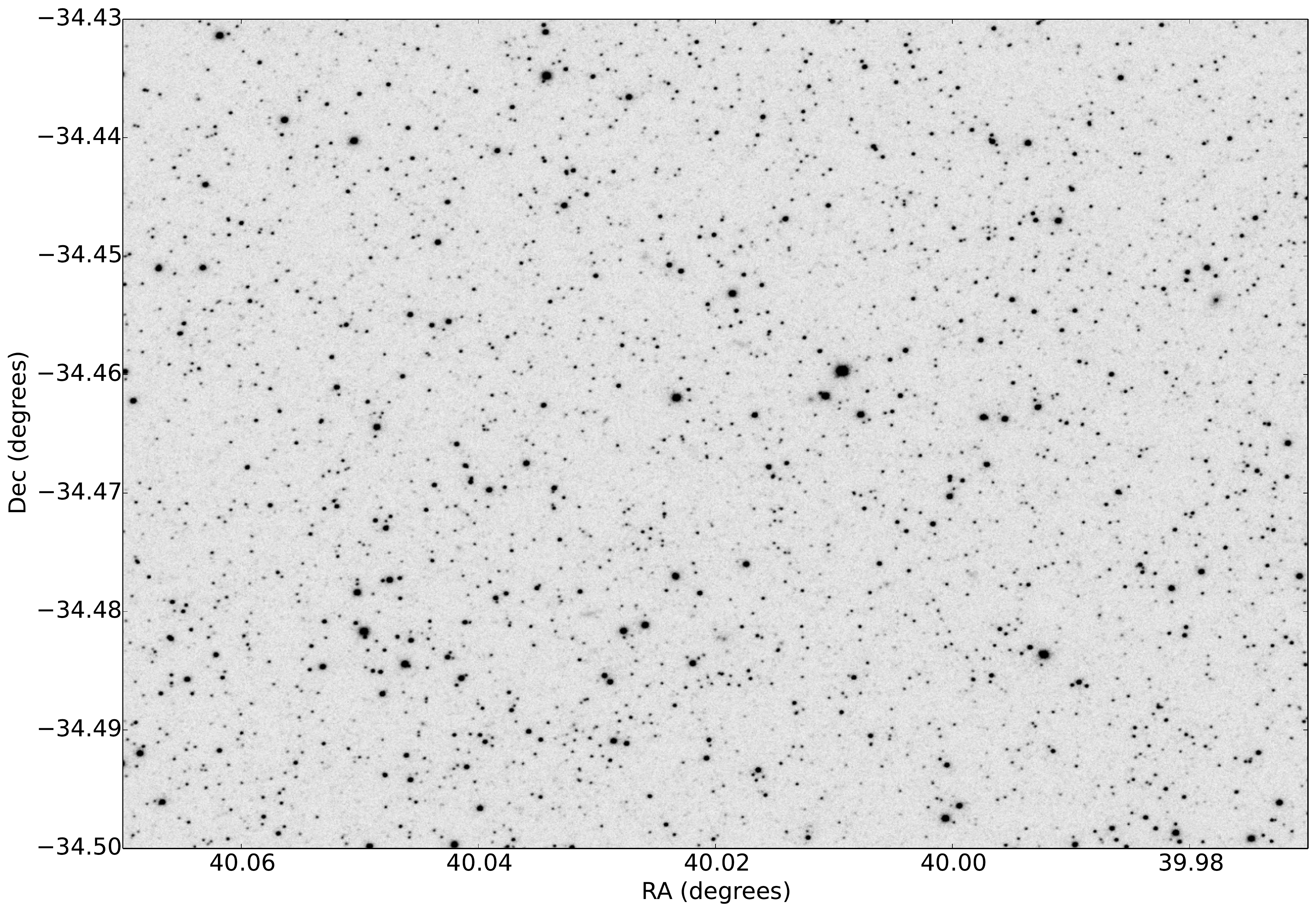}
  }
  \end{center}
  \caption{Panel (a): a three colour image of the whole outer shell region identified in \citet{Coleman2005}. Panel (b): a zoom in on an example portion of the outer shell region, displaying an overdensity of galaxies. Panel (c): a field centred on the Fornax dwarf spheroidal, for comparison. The location of the (b) and (c) fields are shown in Figure \ref{fig:shells}.}
  \label{fig:VST_image_zoom}
\end{figure}

Fornax's strong population gradient (e.g. \citealt{Battaglia2006}; \citealt{Held2010}; \citealt{Cesetti2011a}), with young stars centrally concentrated in the core, is not uncommon for dwarf spheroidals \citep{Harbeck2001}. Such gradients have been detected in the Local Group dwarf spheroidal population via imaging (e.g. \citealt{Harbeck2001}) and spectroscopy (e.g. \citealt{Tolstoy2004}). Similarly, a clear metallicity gradient is seen in Fornax (e.g. \citealt{Walker2009}; \citealt{DeBoer2012}). \citet{Cesetti2011a} detected bimodality in the subgiant branch, and \citet{Battaglia2006} found a hint of splitting near the tip of the RGB. On the other hand, \citet{DelPino2013} and \citet{DeBoer2012} find no clear evidence of splitting.

In addition, Fornax is known to have distinct kinematic populations (e.g. \citealt{Tolstoy2004}; \citealt{Battaglia2006}; \citealt{Walker2011}; \citealt{Battaglia2011}), with the lower velocity dispersion population being characterized by a lower metallicity. The major axes of these two populations are also offset by 30-40$^{\circ}$ (\citealt{Stetson1998}; \citealt{Battaglia2006}). This all points to a very complicated star formation history.

The photometric studies discussed above have tended to focus on strategically placed sub-fields (e.g. \citealt{DeBoer2012}; \citealt{Olszewski2006}), or on coverage extending roughly to the Fornax tidal radius (e.g. \citealt{Battaglia2006}; \citealt{Coleman2008}). In this paper, we use data taken on the 2.6-metre VLT Survey Telescope (VST) as part of the VST ATLAS Survey to explore a 25 square degree field centred on Fornax. This area extends well beyond the nominal tidal radius, and beyond the `outer shell' fields in \citet{Coleman2005} and \citet{Coleman2008}. It offers the perfect opportunity to cleanly explore the spatial distribution of various Fornax stellar populations.

In Section \ref{sec:data} we discuss the VST data used throughout this paper, including the first re-analysis of the region surrounding the purported outer shell in \citet{Coleman2005}, and discussions of colour magnitude diagrams and surface density profiles. In Section \ref{sec:distribution} we use a matched filtering technique to explore the spatial distribution of four populations: two distinct red giant branches populations revealed in our data, as well as the horizontal branch and young main sequence. We use these blue plume stars to explore the inner portions of Fornax for overdensities in Section \ref{sec:innershells}, and discuss the implications of these measurements in Section \ref{sec:discussion}. We conclude in Section \ref{sec:conclusion}. Throughout this paper, colour maps are generated using the `cubehelix' scheme \citep{Green2011}.

\section{Data}
\label{sec:data}

\begin{table*}
\centering
   \begin{minipage}{160mm}
  \caption{Fornax VST ATLAS catalogue} \label{tab:data}
  \begin{tabular}{lccccrccrccr} 
  \hline
  No. & RA (ICRS) & Dec (ICRS) & $g$ & $\Delta g$ & $g$ class & $r$ & $\Delta r$ & $r$ class & $i$ & $\Delta i$ & $i$ class \\
  & ($^\rmn{h}$~$^\rmn{m}$~$^\rmn{s}$) & ($\degr$~$\arcmin$~$\arcsec$) & (mag) & (mag) & &  (mag) & (mag) & & (mag) & (mag) & \\
 \hline
    527 & 2 36 10.7473 & $-31$ 59 56.741 &  15.466 & 0.001 & $-1$   & 14.959 & 0.001 & $-1$ &   14.534 & 0.001 & $-1$ \\
    528 & 2 33 20.4378 & $-31$ 59 56.808  & 22.050 & 0.073 &  $1$ &   21.200 & 0.062 &  $1$  &  20.719 & 0.082  & $1$ \\
    529 & 2 34 22.3833 & $-31$ 59 56.937  & 21.597 & 0.050 & $-1$  &  20.164 &  0.026 & $-1$  &  19.108 & 0.022 & $-1$ \\
    530 & 2 50 37.3408 & $-31$ 59 57.123  & 20.994 & 0.023 & $-1$ &   19.371 &  0.013 & $-1$  &  18.320 & 0.011 & $-1$ \\
    531 & 2 32 19.9185 & $-31$ 59 57.145  & 19.285 & 0.009 & $-1$  &  18.003 & 0.005 & $-1$   & 17.291 & 0.006 & $-1$ \\
    532 & 2 48 10.9424 & $-31$ 59 57.187  & 22.725 & 0.102 &  $1$ &   21.612 & 0.085 &  $1$  &  20.847 & 0.098 & $-3$ \\
    533 & 2 36 34.4322 & $-31$ 59 57.199  & 22.845 & 0.144 & $-3$  &  21.311 & 0.071 & $-3$  &  20.477 & 0.068 & $-1$ \\
    534 & 2 31 56.0799 & $-31$ 59 57.236  & 21.096 & 0.033 & $-1$  &  19.506 & 0.016 & $-1$  &  17.883 & 0.009 & $-1$ \\ 
    535 & 2 28 18.9850 & $-31$ 59 57.308  &  0.000 & 0.000  & $0$  &  21.313 & 0.068 &  $1$  &  20.315 & 0.061 &  $1$ \\
    536 & 2 35 19.1191 & $-31$ 59 57.319  & 20.389 & 0.019  & $1$  &  19.052 & 0.011 &  $1$  &  18.286 & 0.011 &  $1$ \\
  \hline
\end{tabular}

\medskip
A sample of 10 entries in the final Fornax catalogue. The full catalogue is available as online supplementary material. Cambridge Astronomical Survey Unit pipeline classifications are described in Section \ref{sec:data}.

 \end{minipage}
\end{table*}

\begin{figure*}	
  \includegraphics[width=150mm]{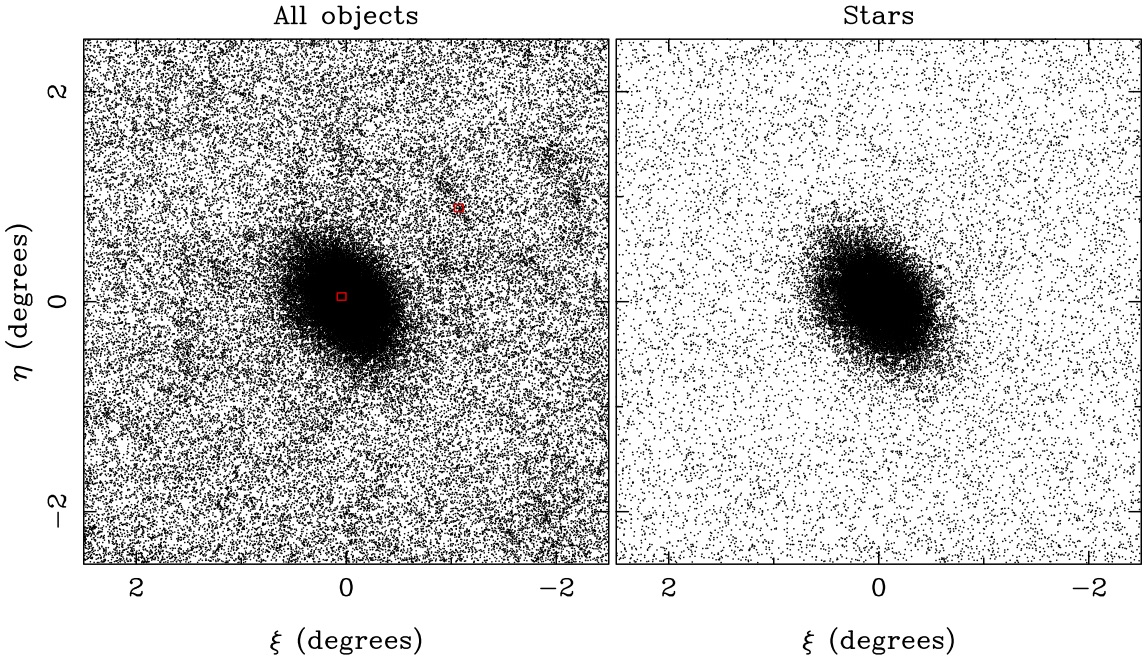}
  \caption{The spatial distribution of objects lying on the red giant branch and red clump locus in the Fornax field. Left panel: all catalogue detections. Right panel: all objects reliably classified by the CASU pipeline as stars. The `outer shell' identified in \citet{Coleman2005} is located $\sim1.4\degr$ north-west of the centre of Fornax in the left panel. It does not appear in the right panel. Red boxes mark the location of the black and white zoom fields provided in Figure \ref{fig:VST_image_zoom}.}
  \label{fig:shells}
\end{figure*}

The data presented in this paper were obtained as part of the ATLAS survey, one of three ESO public surveys currently being carried out using the 2.6-m VLT Survey Telescope (VST). Details of the VST ATLAS survey and Data Release 1 (DR1) can be found in \citet{Shanks13}, \citet{Koposov14} and \citet{Shanks15}. The survey aims to cover an area of several thousand square degrees in the southern celestial hemisphere, in $ugriz$ filters to depths comparable to the Sloan Digital Sky Survey (SDSS) in the north. The median seeing in DR1 is $<1\arcsec$, and the median limiting magnitudes in each band corresponding to $5\sigma$ detection limits are approximately 21.0, 23.1, 22.4, 21.4, 20.2.

The raw data are processed by the Cambridge Astronomical Survey Unit (CASU; \citealt{Irwin2001}). Objects are parameterised and classified morphologically, and catalogues are generated. The ATLAS DR1 photometry used throughout this paper is in Vega magnitudes. These magnitudes were corrected for Galactic extinction using \citet{Schlegel98} dust maps in conjunction with SDSS filter extinction coefficients obtained from \citet{Schlafly11}.

Table \ref{tab:data} contains a sample of 10 catalogue entries. The full catalogue is available as online supplementary material. For each catalogue object we provide ICRS coordinates, $gri$ photometry and associated root-mean-square errors, and CASU pipeline classification in each of the three bands. The pipeline classifications are: 1 = galaxy; 0 = noise-like; $-1$ or $-2$ = reliable star; $-3$ = star/compact galaxy; $-6$ = low average confidence value; $-7$ = contains bad pixels; $-8$ = greater than $1\arcsec$ matching error (matching criteria search radius is $2.5\arcsec$); $-9$ = saturated. These supplementary data are supplied without correction for Galactic extinction. 

\subsection{The Fornax outer shell}
\label{sec:raw_maps}

\begin{table*}
\centering
   \begin{minipage}{160mm}
  \caption{Fornax strutural parameters} \label{tab:density_fits}
  \begin{tabular}{lccccc} 
  \hline
  Parameter & King & King & Sersic & Sersic & \citet{Battaglia2006} \\
  &  & (corrected)  & & (corrected) &  \\
 \hline
 $\alpha_{2000}$ & $2^\rmn{h}39^\rmn{m}52^\rmn{s}$ & $2^\rmn{h}39^\rmn{m}51^\rmn{s}$ & $2^\rmn{h}39^\rmn{m}52^\rmn{s}$ &  $2^\rmn{h}39^\rmn{m}52^\rmn{s}$ & $2^\rmn{h}39^\rmn{m}52^\rmn{s}$ \\
 $\delta_{2000}$ & $-34\degr30\arcmin39\arcsec$ & $-34\degr30\arcmin39\arcsec$ & $-34\degr30\arcmin38\arcsec$ & $-34\degr30\arcmin38\arcsec$ & $-34\degr30\arcmin49\arcsec$ \\
 Ellipticity & $0.31\pm0.01$ & $0.31\pm0.01$  & $0.31\pm0.01$ & $0.31\pm0.01$ & $0.30\pm0.01$ \\
 Position angle & $41.6\degr \pm 0.2\degr$ & $41.5\degr \pm 0.2\degr$ & $41.5\degr \pm 0.2\degr$ & $41.6\degr \pm 0.2\degr$ & $46.8\degr \pm 1.6\degr$ \\
 $r_c$ & $15.4\arcmin \pm 0.1\arcmin$ & $14.6\arcmin \pm 0.1\arcmin$ & -- & -- & $17.6\arcmin\pm0.2\arcmin$ \\
 $r_t$ & $69.5\arcmin \pm 0.4\arcmin$ & $69.7\arcmin \pm 0.3\arcmin$ & -- & -- & $69.1\arcmin\pm0.4\arcmin$ \\
 $r_s$ & -- & -- &  $15.2\arcmin \pm 0.1\arcmin$ & $14.5\arcmin \pm 0.1\arcmin$ & $17.3\arcmin\pm0.2\arcmin$ \\
 Sersic index $m$ & -- & -- & $0.760\pm0.004$ & $0.779\pm0.005$ & $0.71\pm0.01$ \\
 Contamination surface density & $0.501\pm0.001$ & $0.501\pm0.001$ & $0.499\pm0.001$ & $0.499\pm0.001$ & $0.78\pm0.02$ \\
 \quad (stars arcmin$^{-2}$) & & & & \\
 \hline
\end{tabular}

\medskip
Structural parameters obtained from fits to the photometric catalogue above the VST ATLAS $5\sigma$ detection limits. In the corrected case, a modification is made to the models to account for crowding in the inner regions of Fornax (see Section \ref{sec:surface_density} for details). 
 \end{minipage}
\end{table*}

In Figure \ref{fig:VST_image} we present a three-colour image constructed from the $g$, $r$ and $i$ filters of a 16 square degree region centred on the Fornax dwarf spheroidal. The coordinates $(\xi,\eta)$ represent a tangent plane projection centred on Fornax. This image was stitched together from 23 separate pointings in the VST/ATLAS Survey. Of particular interest in this wide-field image is the `outer shell' first noted in \citet{Coleman2005}; its position is marked with a blue box to the North-West of the Fornax dwarf. The dwarf spheroidal itself is clearly visible in this image, and there is perhaps some hint of an overdensity in the blue boxed region. In Figure \ref{fig:VST_image_zoom} we zoom in on this region.

The CASU pipeline uses a curve of growth analysis to classify objects as noise detections, galaxies, and probable stars. To reflect the method used by \citet{Coleman2005} in their original detection of the Fornax outer shell, in Figure \ref{fig:shells} (left panel) we plot all catalogue detections that lie on the Fornax red giant branch (RGB) and red clump (RC) locus in the $g$ and $r$ filters (see Figure \ref{fig:RGB_split_cmds} for the relevant colours and magnitudes). The feature identified in \citet{Coleman2005} as a shell associated with Fornax is clearly visible as an extended feature approximately $1.4\degr$ to the North-West of Fornax.

In the right panel of Figure \ref{fig:shells}, we instead plot all objects lying on the same RGB+RC locus that were reliably classified by the CASU pipeline as stars (aperture photometry classifications of $-1$ or $-2$ in $g$, $r$ and $i$, which corresponds to point sources up to $2\sigma$ from the stellar locus). It is immediately obvious that the `outer shell' has disappeared.

We therefore conclude that the Fornax outer shell does not exist; it is in fact a background overdensity of galaxies. This casts doubt on the possibility of a major merger in the past of the Fornax dwarf spheroidal. The implications of this finding will be discussed further in Section \ref{sec:discussion}. 

The result presented in Figure \ref{fig:shells} does not depend on the magnitude cuts chosen for the RGB+RC. The cut chosen in that figure reaches below the red clump, to $g=22.3$ magnitudes. If we instead cut above the red clump, selecting only the red giant branch stars ($g<21.1$ magnitudes), the features in Figure \ref{fig:shells} are preserved. As a final check, we visually inspected the VST ATLAS images taken in the best seeing. An example image is provided in Figure \ref{fig:zoom_b}, where an overdensity of galaxies is clearly visible; for comparison a field centred on Fornax is provided in Figure \ref{fig:zoom_c}. This inspection confirms the CASU pipeline identifications in the vicinity of the `outer shell'. 

For the remainder of our analysis, we will use all objects in the final catalogue reliably classified as stars in all three bands. We note again that these are aperture photometry classifications of $-1$ or $-2$ in $g$, $r$ and $i$, which corresponds to point sources up to $2\sigma$ from the stellar locus.

\begin{figure*}	
  \includegraphics[width=120mm]{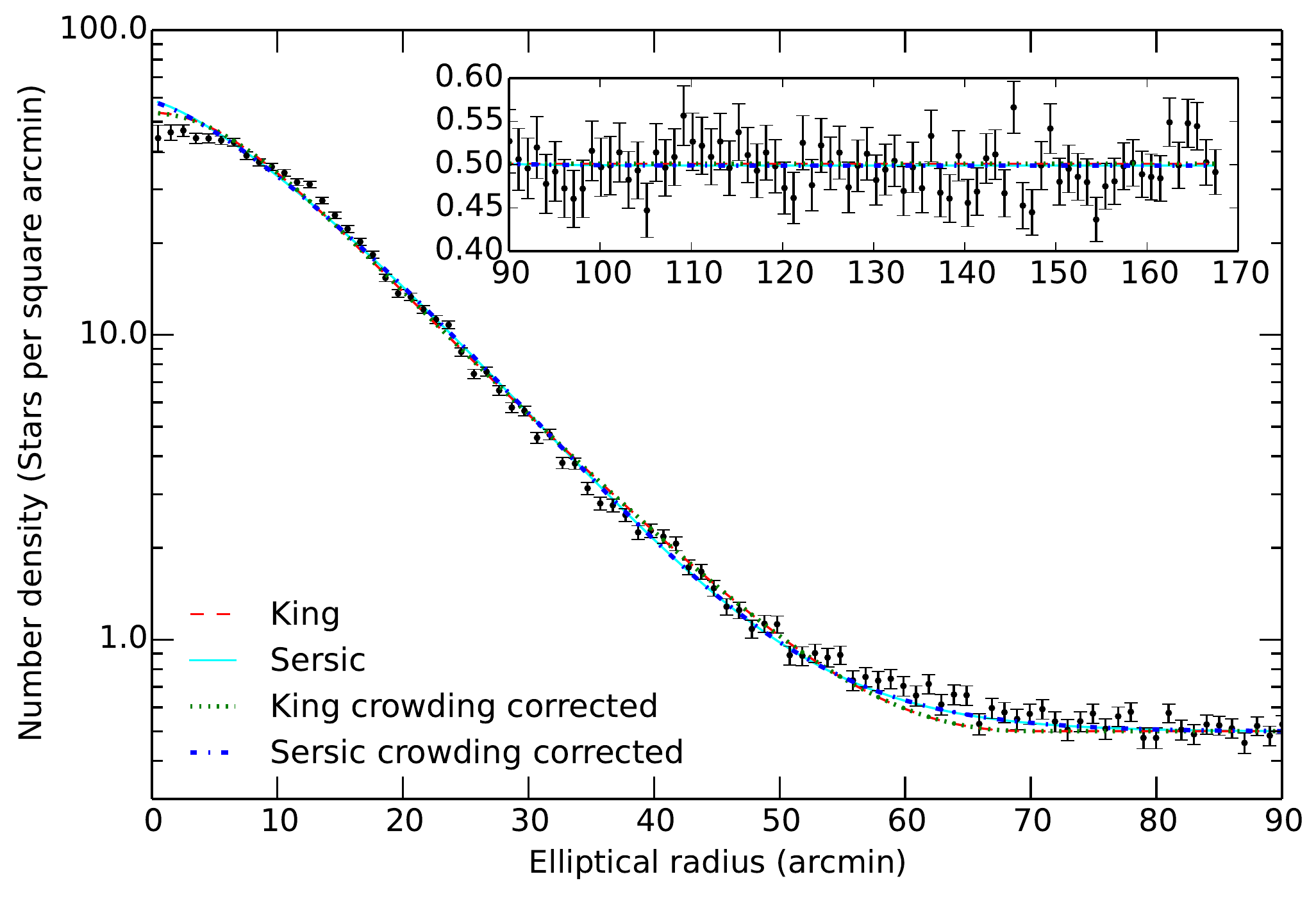}
  \caption{Best King and Sersic profile fits, compared with the data binned in elliptical annuli using the best fit position angle and ellipticity. The fits themselves were performed to the full unbinned dataset above the VST ATLAS $5\sigma$ detection limits, allowing all parameters to vary freely, as described in Section \ref{sec:surface_density}. Fits are provided for models both with and without a correction for crowding in the centre of Fornax, although the correction is minor. Parameters for the fits are provided in Table \ref{tab:density_fits}.}
  \label{fig:profile}
\end{figure*}

\subsection{Surface density profiles}
\label{sec:surface_density}

We fit a surface density profile to the data using an affine-invariant ensemble sampler for Markov chain Monte Carlo (MCMC) of the form proposed by \citet{Goodman2010}. This allowed us to easily sample the entire parameter space, while leaving free the parameters for the centre of Fornax (right ascension and declination), the position angle, the ellipticity, the contamination stellar density, and three parameters for the radial density profile -- a total of 8 parameters. To get a good fit, we ran the MCMC with an ensemble of 50 walkers, each of which did approximately 1000 steps.  

In line with \citet{Battaglia2006}, we fit the Fornax stellar densities using both a King profile \citep{King1962} and a Sersic profile \citep{Sersic1968}. For a detailed explanation of the reasons behind these choices, we refer the reader to \citet{Battaglia2006}. We define the ellipticity $e$ to be $e = 1 - b/a$, where $b$ and $a$ are the semi-minor and semi-major axes of the galaxy respectively. The position angle is taken from North towards East (counter-clockwise, in the figures presented throughout this paper).

Following the method outlined in the appendix of \citet{Richardson2011}, we calculated likelihoods in a way which does not require binning, thus using the full set of information available from the data within the $25$ square degree region around Fornax. The only cuts performed were to select for entries classified as stellar ($-1$ or $-2$) in all three filters, and to exclude entries which fell below the $5\sigma$ detection levels of the VST survey in any of the three filters ($g = 23.1$, $r = 22.4$, $i = 21.4$). The fit was also performed using only the stellar classification cuts, and produced consistent values for all of the parameters with the exception of the contamination stellar density, which was 2 per cent higher in this case.

To account for crowding near the core, we applied the correction detailed in the appendix of \cite{Irwin1984} to the models and repeated all the fits. Again we found that most parameters remained consistent for all fits, with a 2 per cent increase in the contamination in the absence of any magnitude cuts. For the King profile, applying the crowding correction results in a slightly smaller core radius $r_c$. In the Sersic case, the crowding correction leads to a correspondingly smaller scale radius $r_s$ and a $2\sigma$ increase in the Sersic index $m$. This is unsurprising, since the crowding correction effectively increases the number of stars in the Fornax core.

We present relevant parameters from the King and Sersic fits in Table \ref{tab:density_fits}, where the confidence intervals represent the error in the fitting procedure. The results for the appropriate models from \citet{Battaglia2006} are also provided for comparison. Our results are broadly consistent with that work, whose coverage did not extend uniformly beyond the Fornax tidal radius. We note in particular that there is a $\sim5\degr$ offset between our measurements of position angle. Our results are consistent with the earlier work of \citet{Irwin1995}, rather than the figures quoted in \citet{Mateo98} and \citet{Battaglia2006}. We also find slightly smaller King core radius and Sersic scale radius than \citet{Battaglia2006}.

For visual comparison, Figure \ref{fig:profile} shows the best fit models for each profile compared to the data binned in elliptical annuli. These fits were obtained by applying magnitude cuts at the VST ATLAS $5\sigma$ detection limits, and are shown both with and without the crowding correction. As noted in \cite{Battaglia2006}, it is clear that the Sersic profile provides a much better fit around the tidal radius, whereas the King profile is a slightly better fit at small radii, although the data show a flatter core region than either profile can accommodate. 

\subsection{CMDs}
\label{sec:cmds}

\begin{figure*}	
  \includegraphics[width=65mm, angle=270]{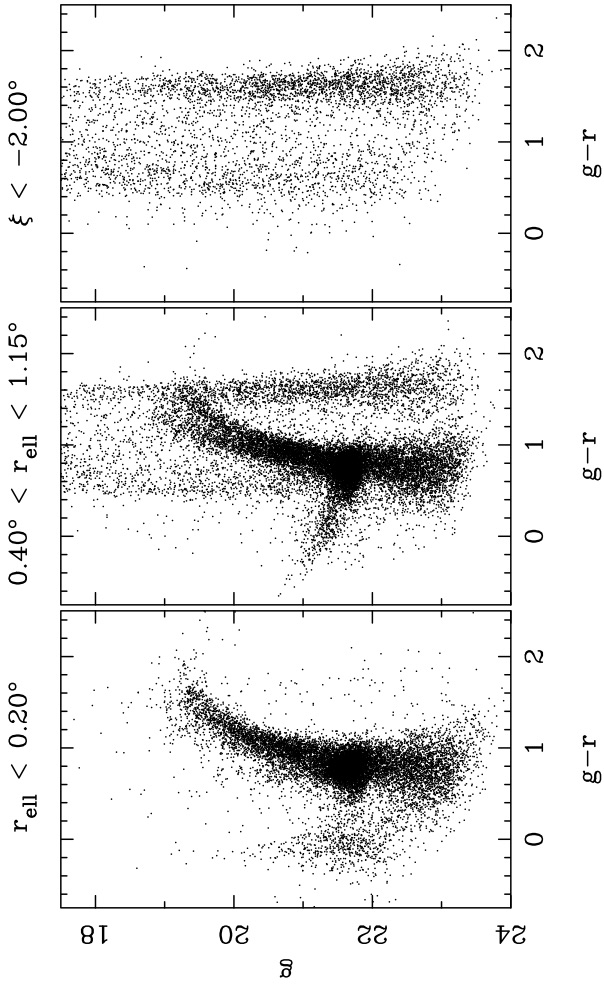}
  \caption{Colour magnitude diagrams in $g$ and $r$ for two regions around Fornax, and a contamination region away from the dwarf. \textit{Left panel:} all stars within an elliptical radius of 0.20 degrees centred on Fornax. \textit{Middle panel:} all stars within an annulus with inner elliptical radius 0.40 degrees, and an outer elliptical radius equal to the Fornax tidal radius ($r_t = 69.1\arcmin$, \citealt{Battaglia2006}). \textit{Right panel:} all stars in a rectangle with $-2.50 < \xi < -2.00$ and $-2.50 < \eta < 2.50$.}
  \label{fig:gr_cmds}
\end{figure*}

\begin{figure*}	
  \includegraphics[width=65mm, angle=270]{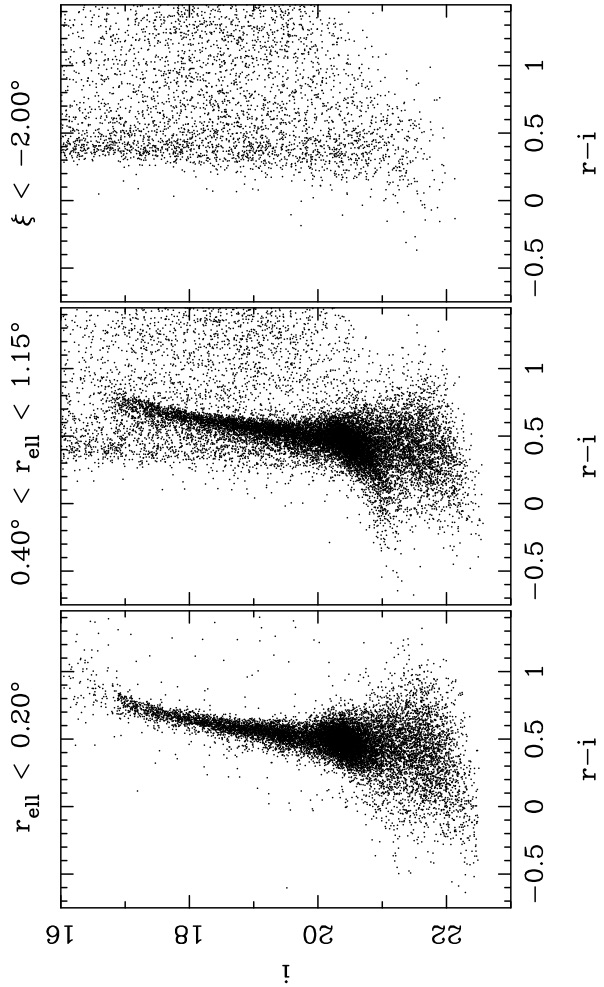}
  \caption{Colour magnitude diagrams in $r$ and $i$, as per Figure \ref{fig:gr_cmds}.}
  \label{fig:ri_cmds}
\end{figure*}

\begin{figure*}	
  \includegraphics[width=65mm, angle=270]{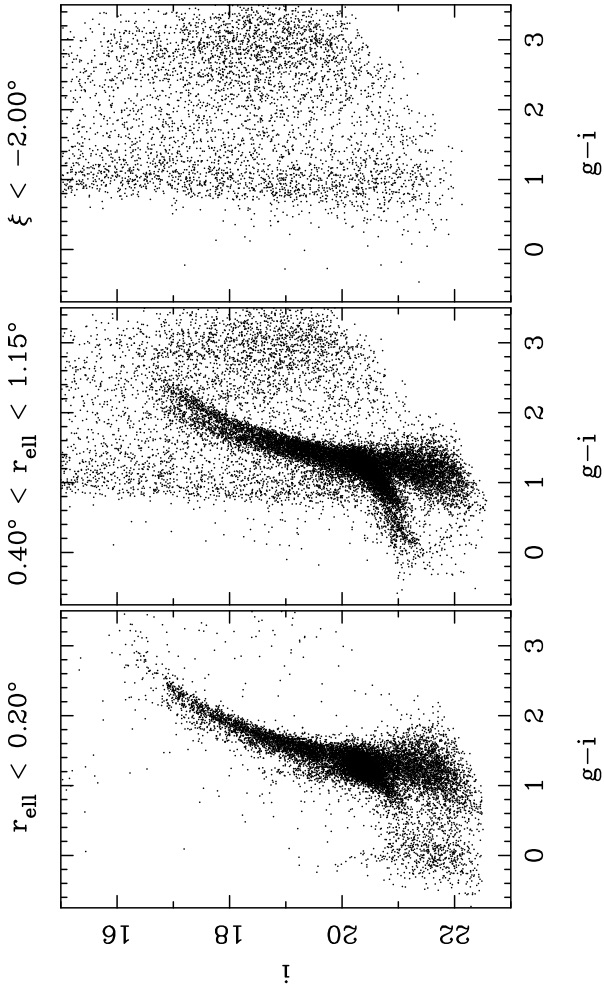}
  \caption{Colour magnitude diagrams in $g$ and $i$, as per Figure \ref{fig:gr_cmds}.}
  \label{fig:gi_cmds}
\end{figure*}

In Figures \ref{fig:gr_cmds}, \ref{fig:ri_cmds}, \ref{fig:gi_cmds} we provide colour magnitude diagrams (CMDs) in three filter combinations: $g$ and $r$, $r$ and $i$, $g$ and $i$ respectively. Each figure shows three CMDs:  for the central regions of Fornax (elliptical radius $r_{ell} < 0.20\degr$, left panels), an annulus extending to the tidal radius ($0.40\degr < r_{ell} < 1.15\degr$, middle panels), and a contamination field consisting of all data with $\xi < -2.00\degr$ (right panels). The contamination field was chosen to best match the depth of the central Fornax fields. We note that while the spatial area covered by the CMDs in the second and third panel is roughly the same ($2.56$ and $2.50$ square degrees), the first panel, centred on Fornax, corresponds to a much smaller area ($0.09$ square degrees).

\begin{figure*}	
  \includegraphics[width=150mm, angle=270]{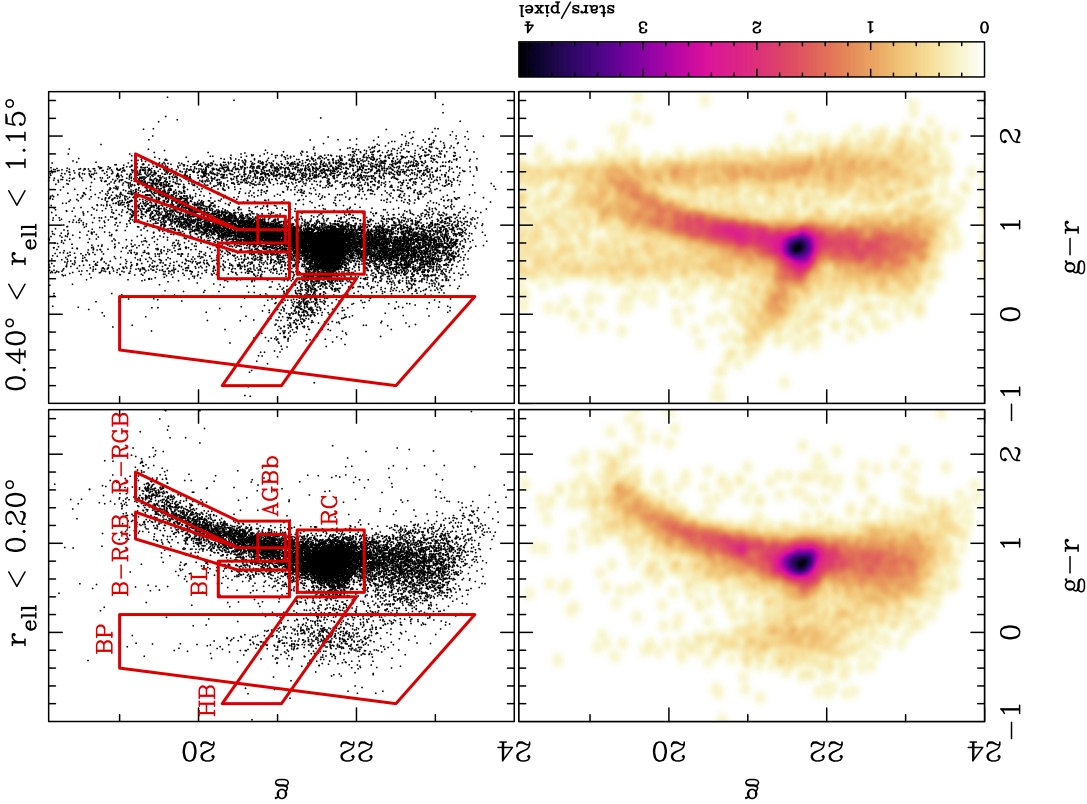}
  \caption{Colour magnitude diagrams for two regions around Fornax. \textit{Left panels:} all stars within an elliptical radius of 0.20 degrees centred on Fornax. \textit{Right panels:} all stars within an annulus with inner elliptical radius 0.40 degrees, and an outer elliptical radius equal to the Fornax tidal radius ($r_t = 69.1\arcmin$, \citealt{Battaglia2006}). A variety of features are marked. HB: horizontal branch, BP: blue plume (a young main sequence, not to be confused with the blue stragglers at the end of the older main sequence turn off), BL: blue loop Helium burning stars, RC: the red clump, AGBb: asymptotic giant branch bump, B-RGB: bluer red giant branch, R-RGB: redder red giant branch (the red giant branch continues below the red clump). In the outer annulus, the metal richer R-RGB population is diminished and a second metal-poorer B-RGB is visible. The bottom row contains Hess diagrams, constructed with 0.01 magnitude pixels and square-root scaling, smoothed using a Gaussian kernel with characteristic radius 0.04 magnitudes. No contamination has been subtracted from these Hess diagrams.}
  \label{fig:RGB_split_cmds}
\end{figure*}

The division into a central region and an outer elliptical annulus was chosen to highlight structure in the red giant branch (RGB). In Figures \ref{fig:gr_cmds} and \ref{fig:gi_cmds} (both $g$-band), the outer CMD shows a clear bifurcation in the RGB. There were hints of this feature in previous work (e.g. \citealt{Battaglia2006}, \citealt{Coleman2008}), however this is the first time it has been revealed with such clarity. Indeed, previous work covering smaller spatial regions (e.g. \citealt{Stetson1998}, \citealt{DeBoer2012}) has favoured continuous star formation at early ages, although \citet{DeBoer2012} note that this could be a consequence of the age resolution in their star formation histories.

Following \citet{Battaglia2006}, we will refer to the two RGBs as the blue RGB (B-RGB) and red RGB (R-RGB). The former is less dense and distributed over a wider area, as expected for an older/more metal poor population. Conversely, although the R-RGB population is visible at all elliptical radii, it is much more centrally concentrated. Young blue plume stars are found only in the innermost regions of Fornax, whereas the horizontal branch is virtually absent in these same regions. These features also point to younger stars being preferentially found in the centre of Fornax, as expected.

In Figure \ref{fig:RGB_split_cmds}, we again provide $g-r$ CMDs for the central region and tidal radius annulus (top panels), along with Hess diagrams of the same (bottom panels). The square root scaling of the colour gradient on the Hess diagrams was again chosen to highlight the bifurcation in the RGB, which is most pronounced in this combination of filters. 

In the top panels of Figure \ref{fig:RGB_split_cmds} we highlight a variety of CMD features, including a blue main sequence of young stars (referred to hereafter as blue plume stars) and an extended blue horizontal branch indicative of an ancient stellar population. These features will be explored in more detail in the following section, wherein we use the Hess diagrams in the bottom panels of Figure \ref{fig:RGB_split_cmds} to construct matched filters.

\section{Spatial distribution}
\label{sec:distribution}

To explore the spatial distribution of the B-RGB, R-RGB, horizontal branch, and blue plume Fornax populations in a statistically robust way, we perform a Poisson-based matched filtering process.  This process was first described in \citet{McMonigal2014}; the salient details are repeated here.

\subsection{Matched filtering technique}

We wish to determine the spatial extent of various populations visible in the Fornax CMDs in Section \ref{sec:cmds}. To do so, we apply a variation on the matched filtering technique of \citet{Kepner1999} and \citet{Rockosi2002}.

Matched filtering in this context is a technique where observed data for an object is used to construct a filter that describes the shape of the population of interest in colour-magnitude space. This filter, in combination with an appropriate contamination, is applied to low-signal regions to pick out the source population from the noise in a statistically meaningful way. The standard matched filtering technique as described in \citet{Rockosi2002} and \citet{Odenkirchen2003} assumes Gaussian statistics, however our CMDs are governed by Poissonian processes. We therefore modify the matched filtering technique as follows.

We describe the number of stars $\lambda_{c,m}$ in a given $(\xi, \eta)$ spatial pixel, with a given colour and magnitude $(c, m)$ with the following model:

\begin{equation}
\lambda_{c,m,\xi,\eta} = \alpha_{\xi, \eta}F_{c, m} + b_{c,m}.
\label{eqn:model}
\end{equation}

$F_{c,m}$ is the foreground filter, in the form of a probability density function in colour-magnitude space. The contamination is described by $b_{c,m}$, and $\alpha_{\xi, \eta}$ is the number of source stars in the spatial pixel $(\xi, \eta)$. It is this last, $\alpha_{\xi, \eta}$, that we wish to determine.

The algorithm we use to determine $\alpha_{\xi, \eta}$ is:

\begin{enumerate}
\item For a given spatial pixel $(\xi,\eta)$, construct an observed Hess diagram.
\item Set $\alpha_{\xi, \eta}=0$.
\item Calculate $\lambda_{c,m,\xi,\eta}$ according to Equation \ref{eqn:model}.
\item Determine the probability for the observed number of stars given $\alpha_{\xi, \eta}$ in each colour-magnitude pixel.
\item Calculate the probability for $\alpha_{\xi, \eta}$ over the entire colour-magnitude space.
\item Increment $\alpha_{\xi, \eta}$ by 1, and return to step (ii) until a maximum $\alpha_{\xi, \eta}$ is reached ($\sim1000$).
\end{enumerate}

We calculate the probability for the observed number of stars $n_{c,m,\xi,\eta}$ given $\alpha_{\xi,\eta}$ using a Poisson distribution:
\begin{equation}
p(n_{c,m,\xi,\eta} | \alpha_{\xi,\eta}) = \exp(-\lambda_{c,m,\xi,\eta})\frac{\lambda_{c,m,\xi,\eta}^{n}}{n!}.
\end{equation}

The probability for $\alpha_{\xi,\eta}$ in a given spatial pixel is the product of the individual probabilities at each pixel in colour magnitude space:
\begin{equation}
p(\alpha_{\xi,\eta}) = \prod{p(n_{c,m,\xi,\eta} | \alpha_{\xi,\eta})}.
\end{equation}
We take the maximum likelihood value of $\alpha_{\xi,\eta}$, around which the probability distributions are sharply peaked, to be the number of source stars in a given spatial pixel. 

\subsection{Filters}
\label{sec:filters}

\begin{figure}
  \includegraphics[width=100mm, angle=270]{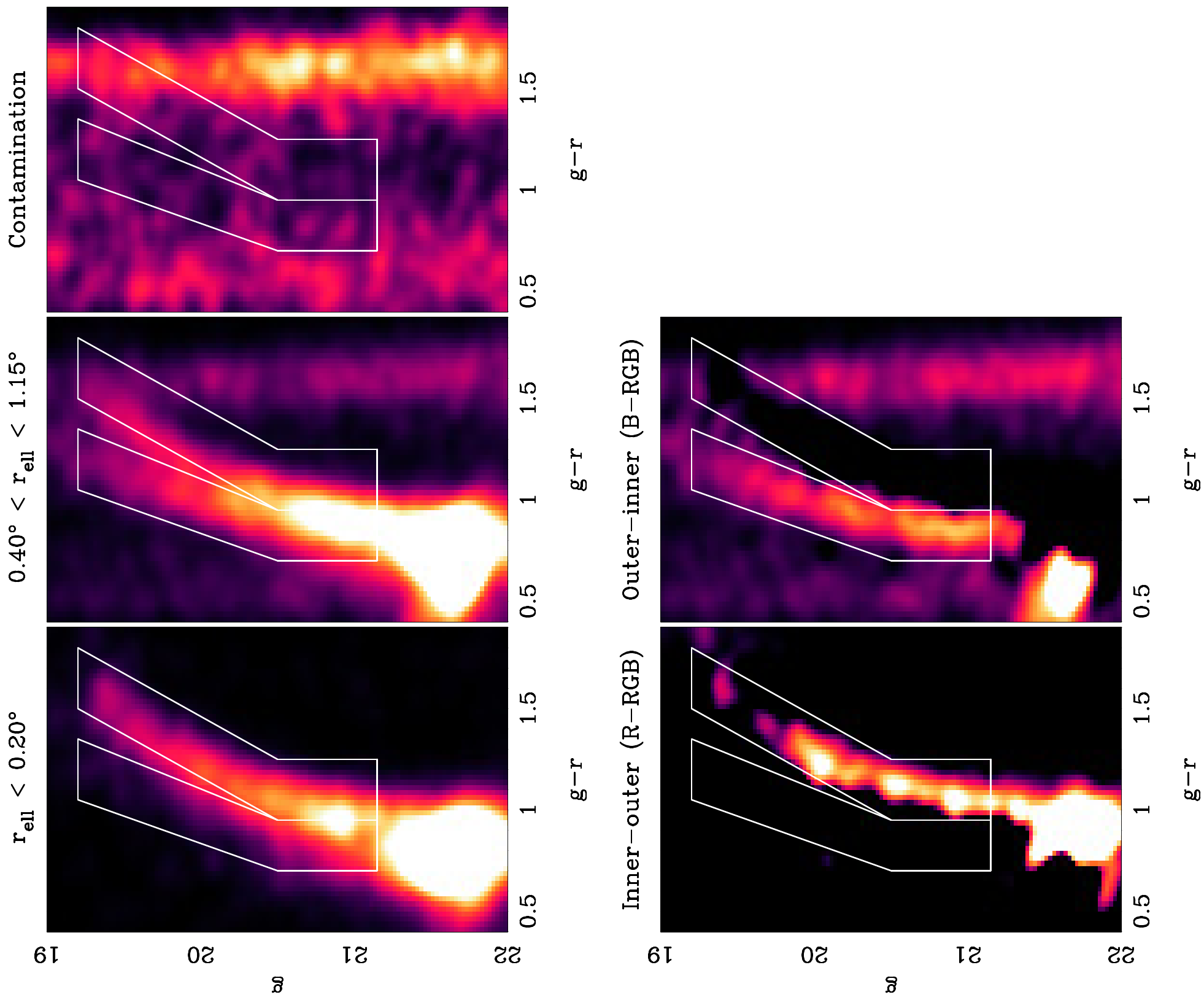}
  \caption{Hess diagrams illustrating the process used to determine appropriate regions to use for red RGB (R-RGB) and blue RGB (B-RGB) filters. Top panels show Hess diagrams constructed from the innermost region of Fornax (\textit{top left}), an outer elliptical annulus around Fornax (\textit{top middle}), and a contamination region north of Fornax (\textit{top right}). Filters were generated using these Hess diagrams. Bottom panels show subtractions of appropriately scaled Hess diagrams to aid in the selection of the filter regions, marked in white. We note that the bottom panels are illustrative only; the filters themselves were constructed from the original Hess diagrams in the top panels.}
  \label{fig:filters}
\end{figure}

\begin{figure*}
  \includegraphics[width=150mm]{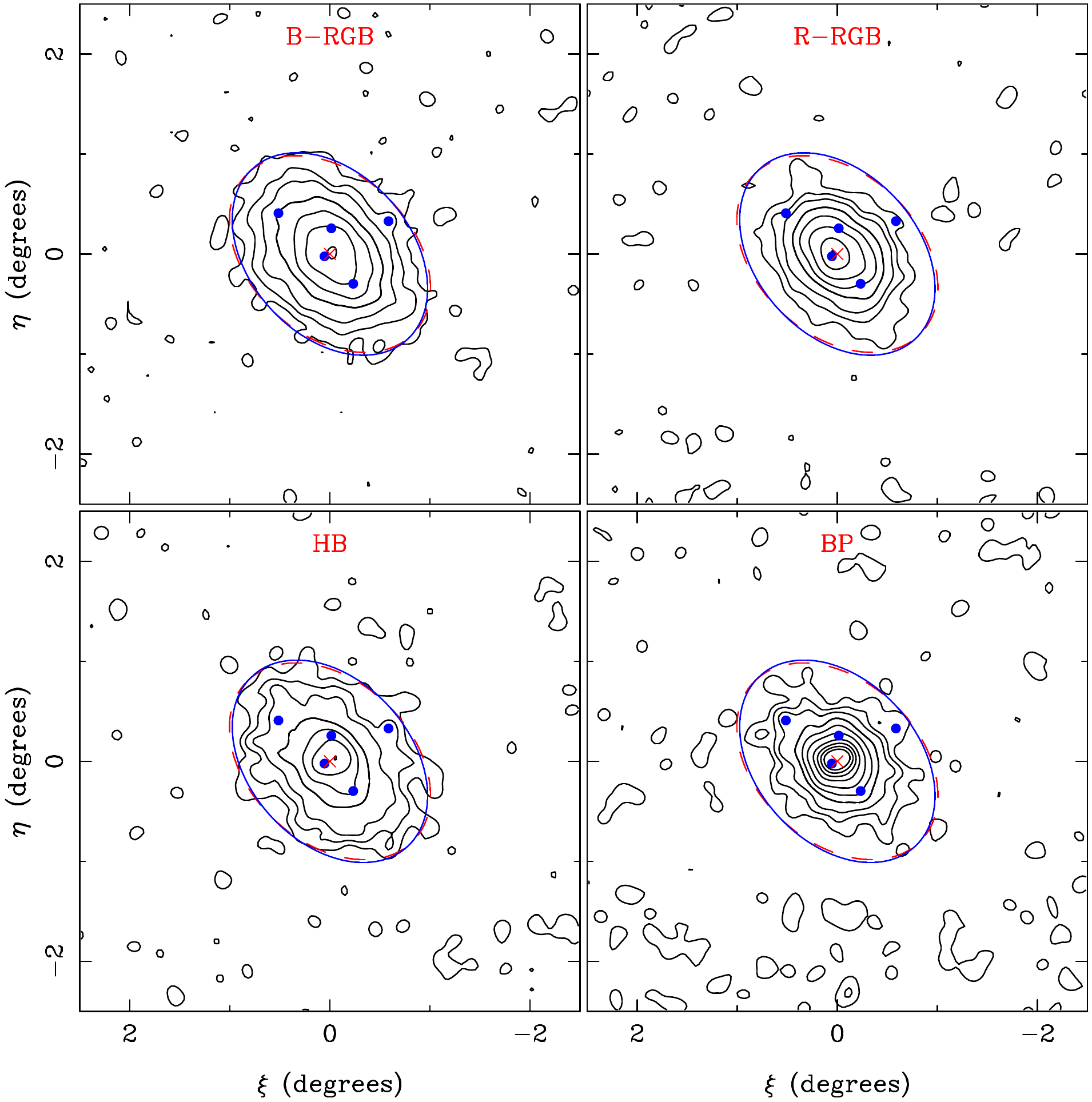}
  \caption{Top left panel: matched filtered stellar density contours for the blue red giant branch (B-RGB) population. Top right panel: the red red giant branch (R-RGB). Bottom left panel: the horizontal branch (HB). Bottom right panel: the young main sequence (blue plume, BP). Pixels are $2\arcmin\times2\arcmin$, smoothed using a Gaussian kernel with $4\arcmin$ dispersion. Contours are displayed at 2, 5, 10, 25, 50, 100, 200, 300 times the (smoothed) RMS value calculated in the contamination region (see text). The tidal ellipse from \citet{Battaglia2006} is marked with a dashed red line, and from this paper with a solid blue line. The centroid (red cross) is marked at the parameters determined in this paper. The known Fornax globular clusters are marked with filled blue circles.}
  \label{fig:RGB_maps}
\end{figure*}

Filters were generated from ($g-r$, $g$) Hess diagrams, as displayed in Figure \ref{fig:RGB_split_cmds} and the top panels of Figure \ref{fig:filters}. This combination of filters was found to best discriminate between the two populations in the split RGB. We repeated the matched filtering process for the ($g-i$, $i$) case, and found comparable results.

Hess diagrams were constructed for three spatial regions in the data: the innermost region ($r_{ell} < 0.20\degr$), an annulus out to the Fornax tidal radius ($0.40\degr < r_{ell} < 1.15\degr$), and a contamination field away from the Fornax population ($\xi < -2.00\degr$). These Hess diagrams consisted of $0.025\times0.025$ mag pixels in both colour and magnitude, smoothed using a Gaussian with 2 pixel dispersion. This was comparable to the observational errors.

The filters were built by truncating the appropriate Hess diagram at appropriate CMD boxes. Figure \ref{fig:filters} illustrates how these boxes were chosen for the B-RGB and R-RGB populations. By using scaled subtractions of the inner ellipse and outer annulus Hess diagrams (bottom panels), we were able to select regions where one population or the other dominated. The final R-RGB and blue plume filters were built from the (raw) inner Hess diagram, and the final B-RGB horizontal branch filters were built from the (raw) outer annulus Hess diagram. The regions selected for the horizontal branch and blue plume populations are illustrated in Figure \ref{fig:bp_cmds}.

\begin{figure*}	
  \includegraphics[width=150mm]{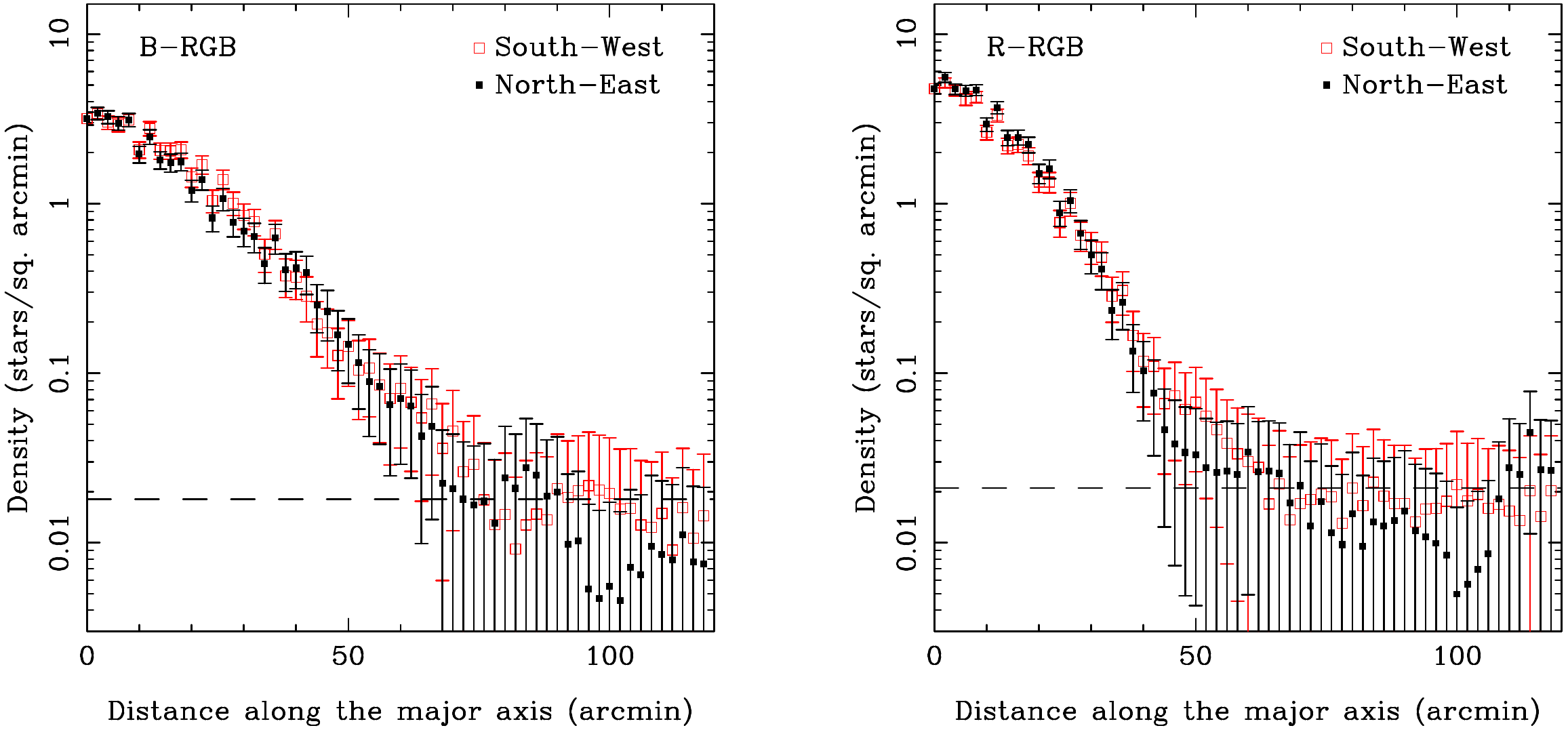}
  \caption{Stellar number count profiles for the blue red giant branch (B-RGB, left panel) and red red giant branch (R-RGB, right panel) Fornax populations, as defined in Section \ref{sec:filters}. These profiles were obtained from matched filtered maps, in a $20\arcmin$ slice along the major axis. The dashed lines mark the RMS stellar density in the contamination region of each map. These RMS values were used to determine the contours displayed in Figure \ref{fig:RGB_maps}.}
  \label{fig:rgb_radial}
\end{figure*}

Foreground filters are probability density functions, and so once a region has been selected from a Hess diagram, they are normalised independently so the area underneath the filter is equal to 1. Conversely, the contamination filter represents a number of stars that is expected in each colour and magnitude pixel, and so it is simply normalised to unit area.

We attempted to push as far down the red giant branch as possible, to maximise our signal while hopefully avoiding contamination from the red clump. As a test, we repeated our matched filtering analysis using filters truncated at $g = 20.5$ (i.e. removing the vertical section of the selection regions). Although the signal was slightly diminished in the resulting matched filtered maps, the structure of the maps was essentially unchanged.

\subsection{Matched filtered maps}

\begin{figure}	
  \includegraphics[width=60mm, angle=270]{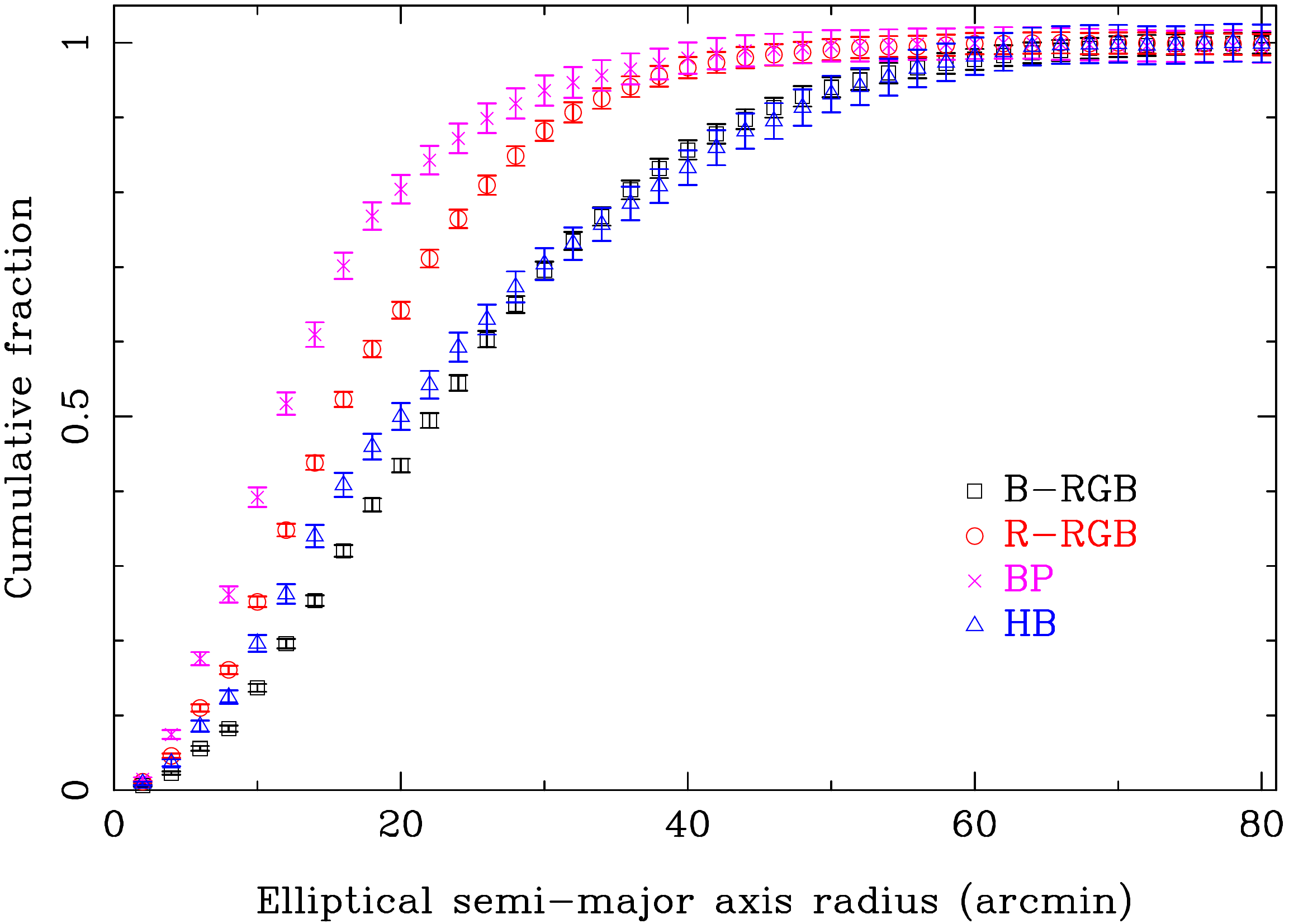}
  \caption{The cumulative fraction of stars as a function of elliptical semi-major axis radius for four populations: the blue red giant branch (B-RGB, black squares), the red red giant branch (R-RGB, red circles), the blue plume (BP, purple crosses), and the horizontal branch (HB, blue triangles). Number counts were obtained from matched filtered maps, accounting for the expected RMS contamination in each ellipse. Each population is normalised independently. We note that the B-RGB and HB follow approximately the same profile, and so are likely to be coeval.}
  \label{fig:cumulative}
\end{figure}

\begin{figure*}	
  \includegraphics[width=170mm]{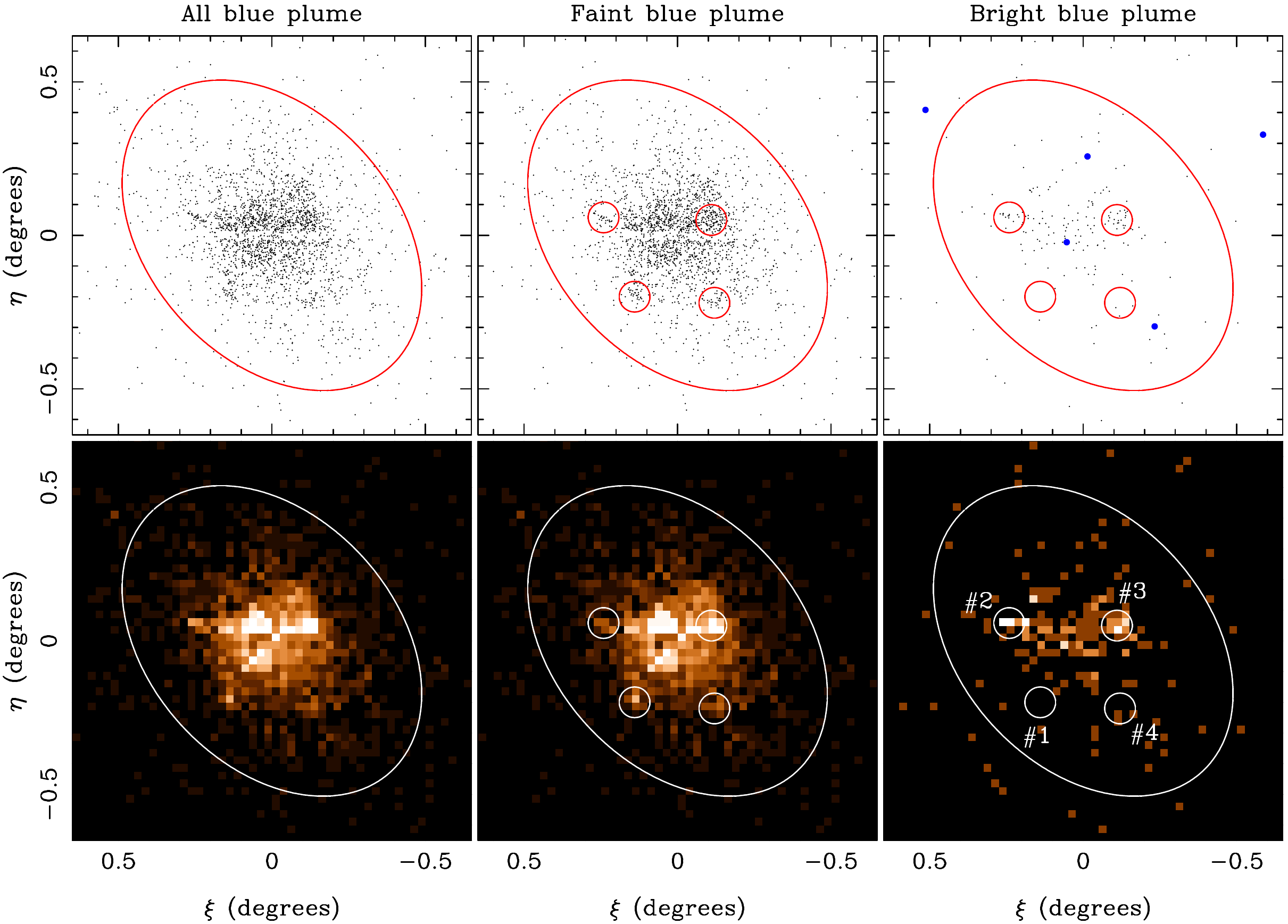}
  \caption{The spatial distribution of Fornax blue plume stars, selected from the $g$ and $r$ bands. Top panels show individual catalogue objects. Bottom panels, the stars have been binned into $1.5\arcmin\times1.5\arcmin$ pixels, and a linear colour gradient chosen to highlight overdensities. Left panels: the entire blue plume population (excluding the horizontal branch). Middle panels: blue plume fainter than the horizontal branch. Right panels: blue plume brighter than the horizontal branch. In each panel, an ellipse is drawn with Fornax's ellipticity and position angle (this paper), at half its measured tidal radius. Filled blue circles in the top right panel mark the locations of known globular clusters. Four overdensities are highlighted with $3\arcmin$ diameter circles, and labelled in the bottom right panel. Overdensity 1 was first detected in \citet{Coleman2004}, overdensity 2 in \citet{DeBoer2013}. Overdensities 3 and 4 are new. CMDs for these over dense regions can be found in Figure \ref{fig:bp_cmds}.}
  \label{fig:bp_maps}
\end{figure*}

\begin{figure*}	
  \includegraphics[width=130mm]{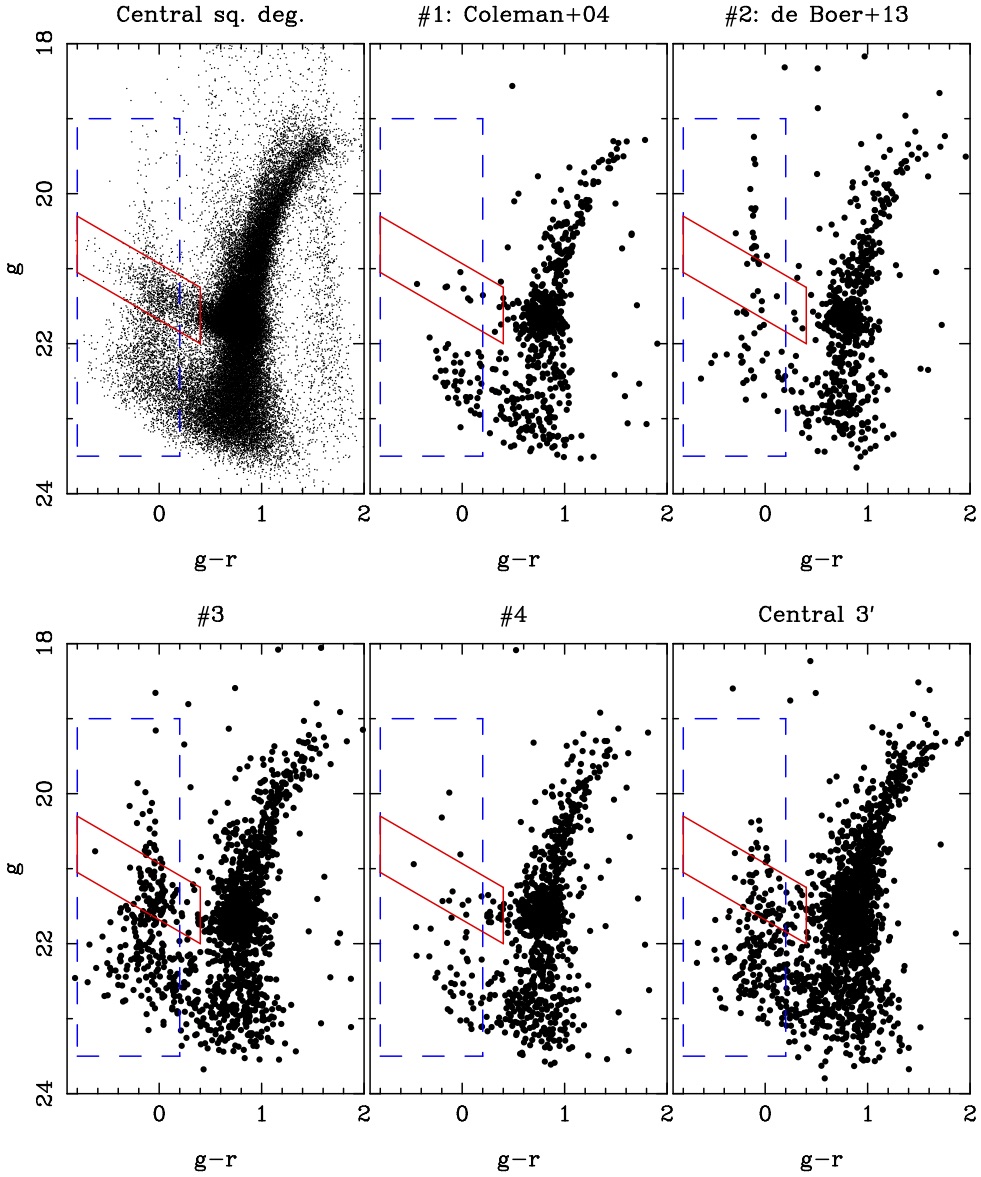}
  \caption{CMDs for overdensities in the Fornax field, following the labelling convention in Figure \ref{fig:bp_maps}. Top left: the full CMD for the central square degree around Fornax. The blue dashed box marks the blue plume (BP), and the solid red box the horizontal branch (HB). The HB divides the blue plume into two regions: the Brighter BP and the Fainter BP. Top middle: a $3\arcmin$ radius around the overdensity identified in \citet{Coleman2004}. Top right: a $3\arcmin$ radius around the overdensity identified in \citet{DeBoer2013}. Bottom left: a $3\arcmin$ radius around the overdensity in Figure \ref{fig:bp_maps} identified in the Brighter BP population. Bottom middle: a $3\arcmin$ radius around the overdensity in Figure \ref{fig:bp_maps} identified in the Fainter BP population. Bottom right: a $3\arcmin$ radius around the Fornax centroid identified in this paper. Not the similarities between overdensities 2 and 3, 1 and 4.}
  \label{fig:bp_cmds}
\end{figure*}

Matched filtered maps for four populations are provided in Figure \ref{fig:RGB_maps}: the B-RGB (older/more metal-poor stars; top left panel), R-RGB (younger/more metal-rich stars, top right panel), horizontal branch (indicative of an ancient stellar population, bottom left panel), and the blue plume (a young main sequence, bottom right panel). We note that the panels are normalised separately, based on their respective root-mean-square contamination in the $\xi < -2.00\degr$ contamination region. This level is 0.072 stars pixel$^{-1}$ for the B-RGB (top left), 0.084 stars pixel$^{-1}$ for the R-RGB (top right), 0.034 stars pixel$^{-1}$ for the horizontal branch (bottom left), and 0.026 stars pixel$^{-1}$ for the blue plume (bottom right). In each panel we also plot the tidal ellipse and centroid determined in this paper, and the five known Fornax globular clusters.

It is clear from these maps the the B-RGB population is significantly more diffuse than the R-RGB population. The R-RGB population is more centrally located, and displays somewhat less regular density contours, particularly towards the outer regions. This suggests somewhat patchy star formation. Neither population shows any clear evidence of extra-tidal features, suggesting that Fornax is not currently undergoing tidal disruption. This is in line with proper motion measurements placing Fornax at a relatively distant perigalacticon of $\sim150$kpc (e.g. \citealt{Dinescu2004}).

The horizontal branch, while consisting of fewer stars than the B-RGB, appears to follow essentially the same spatial distribution as that population. This suggests that the two populations are coeval. The blue plume stars are once again quite centrally located, and their density contours significantly more circular than any of the other populations examined here.

In Figure \ref{fig:rgb_radial}, we provide stellar radial profiles for the red giant branch populations, obtained from the matched filtered maps. These profiles were determined along a $20\arcmin$-wide slice through the matched filtered maps, along the major axis. The left panel shows the B-RGB population, and the right panel the R-RGB population. The latter is denser and more centrally located, and appears slightly asymmetric, extending out to the tidal radius on towards the South-West, but truncating earlier on the North-East.

We also map the cumulative number of stars in each of the four populations, normalised to their peak number, as a function of elliptical semi-major axis radius in $2\arcmin$ increments. These profiles are shown in Figure \ref{fig:cumulative}. They highlight that the blue plume population is the most compact (although we note again that it does not have the same elliptical shape as the other three populations), followed by the metal-richer R-RGB. The metal-poorer B-RGB and the horizontal branch follow essentially the same profile as a function of elliptical radius, suggesting that these two populations are coeval.

\section{Inner overdensities}
\label{sec:innershells}

In Section \ref{sec:raw_maps} we discussed the misclassification of a background overdensity of galaxies as a shell of stars, presumed to have been associated with a significant merger in Fornax's past. Both \citet{Coleman2004} and \citet{DeBoer2013} have reported the detection of stellar overdensities much closer to the centre of Fornax. In the case of \citet{DeBoer2013}, a new overdensity was identified in blue plume (BP) stars. 

Guided by this discovery, we examined the spatial distribution of the blue plume population. CMD selection boxes were built using the central square degree of our $g$ and $r$ data (see Figure \ref{fig:bp_cmds}, top left panel), where the BP population is most easily discriminated. We split the BP into three populations: the entire BP (excluding the horizontal branch), the BP below the horizontal branch (``faint blue plume''), and the BP above the horizontal branch (``bright blue plume''). We were motivated to look at the brighter and fainter BP populations separately by the unusually bright blue plume in the \citet{DeBoer2013} overdensity when compared with the rest of the Fornax population.

In Figure \ref{fig:bp_maps} we present maps of the blue plume stars in these three CMD regions: the full blue plume (left panels), the faint blue plume (middle panels), and the bright blue plume (right panels). In the top panels, the individual stars are plotted, whereas in the bottom panels they are binned in $1.5\arcmin\times1.5\arcmin$ pixels, with a linear colour gradient chosen to highlight regions of higher density. In each panel, an ellipse is plotted with the ellipticity and position angle of Fornax, at an elliptical radius of half the Fornax tidal radius. 

Both the \citet{Coleman2004} and \citet{DeBoer2013} overdensities (circles 1 and 2 respectively) are clearly visible in the full BP population. The \citet{Coleman2004} region particularly stands out in the fainter BP, and the \citet{DeBoer2013} region is a marked overdensity in the brighter BP population. This is in agreement with \citet{DeBoer2013}, who showed that while their new overdensity consisted of unusually young stars, the \citet{Coleman2004} overdensity was much more similar to the standard Fornax population.

The \citet{Coleman2004} feature does appear to have a shell-like morphology, whereas the \citet{DeBoer2013} feature is more radial, although still extended. Chip gaps certainly have an impact on structure we see in the VST ATLAS data, although the binned stellar density plots in Figure \ref{fig:bp_maps} do indicate a patchy distribution of young stars. 

We identify two new regions away from the very centre of Fornax that appear over-dense, one in the brighter BP population (circle 3) and one in the fainter BP population (circle 4). Neither is as significant as the earlier detections. The feature in circle 3, in the bright BP, is more diffuse than any of the other overdensities. The feature in circle 4 does appear to have a somewhat arc-like morphology, although the low number count of stars could well be misleading.

In Figure \ref{fig:bp_cmds} we display CMDs for the central square degree of Fornax, and for $3\arcmin$ circles centred on each of the overdensities in Figure \ref{fig:bp_maps}. The bright blue plume in the \citet{DeBoer2013} overdensity is striking (top right panel). Our new overdensity 3 (bottom left panel) is somewhat similar, suggesting that very recent star formation may be occurring in more than one isolated pocket outside the centre of Fornax. The least significant of the new overdensities that we've highlighted here, number 4 (bottom middle panel), is very similar to the feature first identified in \citet{Coleman2004}.

\section{Discussion}\label{sec:discussion}

The most prominent result of our analysis of wide-field VST ATLAS data of the Fornax dwarf spheroidal concerns the outer shell first reported in \citet{Coleman2005}. With these new data, taken on a larger telescope in better observing conditions, the outer `shell' is revealed to be a background overdensity of galaxies (see Figure \ref{fig:shells}), and not associated with Fornax at all. This putative outer shell had been taken as evidence for a major interaction in Fornax's past, and the gold-star example of dwarf-dwarf interactions in the Local Group (e.g. \citealt{Deason2014}, along with Andromeda II \citealt{Amorisco2014}). 

A key feature of the Fornax merger hypothesis put forward by \citet{Coleman2004} and \citet{Coleman2005} was the presence of shell-like structures on opposite sides of the Fornax minor axis: a smaller inner shell on the southeast, and the much more prominent outer `shell'  to the northwest. These shells were also associated with `lobes', consisting of slight overdensities of stars outside the tidal ellipse along the minor axis. Although the ESO/MPG 2.2m observations reported in \citet{Coleman2008} specifically cover the outer `shell', no information on that feature was reported.

The dwarf merger scenario has inspired theoretical work by \citet{Yozin2012}, in which they use N-body simulations to conclude that the hypothesis is plausible. This result hinged almost entirely on the creation of an outer shell, and further predicted the presence of tidal tails. We find no clear evidence for tidal features in our analysis, in line with previous observational and theoretical studies (see e.g. \citealt{Penarrubia2009b}).

It should be noted that even despite the incorrect identification of the outer `shell', reconciling a merger event with the Fornax star formation histories and the properties of the inner shells has proven problematic. \citet{Olszewski2006} demonstrated that the stars in the inner shell identified by \citet{Coleman2004} were consistent with younger ($\sim1.4$Gyr) Fornax field stars, suggesting that they were the product of pre-enriched gas in Fornax rather than externally accreted.

This result is confirmed in \citet{DeBoer2013}, where the authors conduct a detailed analysis of the CMDs and star formation histories of the inner \citet{Coleman2004} clump. They also find a new clump, which is even younger still (100-200 Myr, the youngest age bin considered in their analysis). By filtering only for bright blue plume stars, we reveal here a third clump of young star formation, offset by $\sim7\arcmin$ from the centre of Fornax. This feature displays a CMD very similar to the \citet{DeBoer2013} overdensity (see Figure \ref{fig:bp_cmds}).

Although it is tempting to discard the merger hypothesis altogether, the situation is complicated somewhat by kinematic data. Using a technique developed by \citet{Walker2011}, \citet{Amorisco2012} presented evidence that a sample of Fornax red giants consists of three distinct subpopulations, each with different kinematic properties. In particular, in two separate datasets (\citealt{Walker2009} and \citealt{Battaglia2006}) they find that two red giant populations (one metal poor and one intermediate metallicity) are rotating in opposite directions. They interpret this as possible evidence for the merger of a bound pair. This counter-rotating result hinges on the proper motion measurements of \citet{Piatek2007}. Clearly, more spectroscopic data are needed, focussing in particular on the more metal poor red giants located in the outskirts of Fornax, where spectroscopic coverage tends to be sparse.

Nevertheless, by demonstrating that the Fornax outer shell does not in fact exist, we cast doubt on the possibility of a significant merging event in the history of Fornax. The causes of the young bursts of star formation and the (possible) counter-rotation in the metal poor and intermediate metallicity red giants are still unclear, although the weight of evidence supports the conclusion that all of these stars were generated from Fornax gas, rather than material accreted from elsewhere.

The wide-field VST ATLAS data that we present also clearly reveals a bifurcation in the Fornax red giant branch. There have been hints of this division in previous work (e.g. \citealt{Battaglia2006}; \citealt{Walker2011}), however it is clearly visible in the VST $(g-r,g)$ CMDs. The clarity of this feature is primarily due to the uniform wide-field coverage of our data. It has long been known that Fornax displays a radial age gradient (e.g. \citealt{Stetson1998}; \citealt{Saviane2000}; \citealt{Battaglia2006}; \citealt{DeBoer2012}; \citealt{DelPino2013}).

The star formation histories derived in \citet{DeBoer2012} and \citet{DelPino2013} differ in their details. \citet{DeBoer2012} prefer an intermediate age of 1-10 Gyr for the dominant Fornax population, with a dominant RGB population at $\sim4$Gyr. They do note, however, that their age resolution may smooth out any old features in the star formation history.

In contrast, \citet{DelPino2013} prefer a peak in star formation at approximately 8 Gyr ago in the central regions of Fornax, and 10 Gyr and older in outer regions (located roughly at the Fornax tidal radius). They attribute this difference to shallower depth in the \citet{DeBoer2012} photometry. We note that our uniform, wide-field VST data only clearly reveal the second distinct peak in the RGB when the data are clearly extended to the tidal radius; repeating the analysis of \citet{DelPino2013} all the way out to that radius may well heighten the signal in the older population. It is perhaps significant to note that the \citet{DelPino2013} results, while not ruling out mergers as significant in the evolution of Fornax, are completely consistent with evolution where mergers did not play a significant role.

\section{Conclusion}\label{sec:conclusion}

We present photometric data from the VLT Survey Telescope ATLAS Survey of a 25 square degree region centred on the Fornax dwarf spheroidal. These uniform, wide-field data reach down to depths of 23.1 in $g$, 22.4 in $r$, and 21.4 in $i$, and reveal a variety of interesting features.

Most significantly, we demonstrate that the outer `shell' first reported in \citet{Coleman2005} is in fact not a stellar overdensity, but an overdensity of background galaxies. Most likely the misidentification in the original data was due to the observing conditions. No follow-up has been reported on this feature since its original appearance in the literature, although its presumed existence has been used to support arguments that Fornax underwent a significant merger with another dwarf-sized object in its past.

Multiple lines of evidence are suggestive of some sort of merger activity, including isolated regions of young star formation away from the core of Fornax (\citealt{Coleman2004}; \citealt{DeBoer2013}; this work), and the possible detection of counter-rotation in the metal poor and intermediate metallicity populations \citep{Amorisco2012}. Although our new results do not rule out the merger hypothesis entirely, they certainly remove a line of supporting evidence.

We also detect for the first time a completely unambiguous bifurcation in the red giant branch. This suggests that rather than smooth star formation until a single dominant burst at intermediate ages, Fornax has undergone at least two distinct bursts of significant star formation activity. This result is in line with work by \citet{DelPino2013}, whose analysis focused on stars relatively close to the core of Fornax, and measured two distinct periods of increased star formation, $\sim8$ and $\sim10$ Gyr ago.

Using a matched filtering technique, we have mapped out the two red giant branch populations (called here B-RGB and R-RGB), as well as the horizontal branch (HB) and young main sequence (blue plume, BP). We find that the B-RGB -- most likely corresponding to an older population -- is significantly more spatially diffuse, extending all the way out to the Fornax tidal radius. Conversely, the R-RGB population is more centrally focused, and displays a somewhat less elliptical shape than its B-RGB counterpart. The blue horizontal branch, which indicates an ancient stellar population, follows the same basic profile as the B-RGB. These two populations likely evolved together. Lastly, the blue plume stars are the most centrally located, and their density contours are significantly more circular than the other Fornax populations.

We have also identified an additional overdensity of young stars, $\sim7\arcmin$ from the centre of Fornax. This overdensity is very similar to the new feature detected in \citet{DeBoer2013}, and was selected against the relatively dense central regions of Fornax by focussing on bright blue plume stars (above the horizontal branch).

There is a need to pursue spectroscopic studies of Fornax in more detail, and to considerably larger radii. The intermediate age red giant population dominates in the central region, thus limiting our knowledge of the older population. In particular, by studying stars at these larger radii we can hope to get a firmer handle on the star formation history of Fornax, and determine whether the counter-rotation -- perhaps the strongest surviving argument for a significant merger in Fornax's history -- is real.

\section*{Acknowledgments}
We wish to thank the anonymous reviewer for helpful suggestions which significantly improved this paper. NFB and GFL thank the Australian Research Council (ARC) for support through Discovery Project (DP110100678). GFL also gratefully acknowledges financial support through his ARC Future Fellowship (FT100100268). BM acknowledges the support of an Australian Postgraduate Award. This paper is based on observations obtained as part of the VST ATLAS Survey, ESO Progam, 177.A-3011 (PI: Shanks). The UK STFC is acknowledged for supporting the Cambridge Astronomical Survey Unit. Work at Durham is supported by the Science and Technology Facilities Council grant ST/L00075X/1. NFB would like to thank (or possibly blame) E. N. Taylor for the title of this paper.

\bibliography{nbate_fornax} 
\bsp

\label{lastpage}

\end{document}